\documentclass{article}

\usepackage[nonatbib]{nips_2018}

\usepackage[utf8]{inputenc} % allow utf-8 input
\usepackage[T1]{fontenc}    % use 8-bit T1 fonts
\usepackage{hyperref}       % hyperlinks
\usepackage{url}            % simple URL typesetting
\usepackage{booktabs}       % professional-quality tables
\usepackage{amsfonts}       % blackboard math symbols
\usepackage{amsmath}
\usepackage{color}
\usepackage{nicefrac}       % compact symbols for 1/2, etc.
\usepackage{microtype}      % microtypography
\usepackage[pdftex]{graphicx}
\usepackage{wrapfig}
\usepackage{lipsum}

\newcommand{\figref}[1]{Figure \ref{#1}}

\newcommand{\secref}[1]{Section \ref{#1}}
\graphicspath{{images/}}

\title{Orchestrate: Infrastructure for Enabling Parallelism during Hyperparameter Optimization}

\author{
  Alexandra Johnson, Michael McCourt \\
  SigOpt\\
  San Francisco, CA, USA \\
  \texttt{\{alexandra, mccourt\}@sigopt.com} \\
}

\begin{document}

\maketitle

\begin{abstract}

    Two key factors dominate the development of effective production grade machine learning models.
    First, it requires a local software implementation and iteration process.
    Second, it requires distributed infrastructure  to efficiently conduct
    training and hyperparameter optimization.
    While modern machine learning frameworks are very effective at the former,
    practitioners are often left building ad hoc frameworks for the latter. 
    We present SigOpt Orchestrate, a library for such simultaneous
    training in a cloud environment.
    We describe the motivating factors
    and resulting design of this library, feedback from initial
    testing, and future goals. 
\end{abstract}

\section{Introduction}
Deep learning models have enjoyed broad adoption \cite{lecun2015deep}
because of the development of popular libraries, such as
MXNet \cite{chen2015mxnet},
PyTorch \cite{paszke2017automatic} and
Tensorflow \cite{Abadi:2016:TSL:3026877.3026899}.
These libraries have provided an efficient and stable framework in
which to develop models.

For these deep learning models to perform well,
meta-decisions regarding their architecture and hyperparameters must be
made; conducting this model tuning efficiently presents a challenge
both in terms of strategy and implementation.  Often times, the strategy
for hyperparameter tuning involves defining some measurement of
generalization quality for a given model and then using a black-box
optimization strategy to find an optimal
configuration \cite{hutter2011sequential}.
Suitable strategies include
grid search \cite{bergstra2011algorithms},
random search \cite{bergstra2012random},
evolutionary strategies \cite{young2015optimizing},
swarm intelligence methods \cite{blum2008swarm},
and Bayesian optimization \cite{frazier2018tutorial, shahriari2016taking}.

Each of these strategies requires training a model many times,
each with different hyperparameters/architecture.
Training several models in parallel can reduce the necessary
wall clock time required to complete this important step.
Running multiple models in a local development environment is likely infeasible, usually because of
the specialized hardware required for deep learning
models.  As such, distributed infrastructure for parallel
model trainings is a necessary
component of an efficient model building pipeline.
Deep learning models also require
a great deal of high quality labeled data,
but this topic is not discussed in this article.

We present a library, Orchestrate, which seeks to manage the
infrastructure complications
fundamental to parallel hyperparameter tuning.
Orchestrate was designed to work with SigOpt, a
cloud-based optimization API for hyperparameter tuning in
parallel \cite{pmlr-v84-martinez-cantin18a}.  The goal of Orchestrate
is to provide the necessary infrastructure to coordinate and
simultaneously execute multiple hyperparameter configurations suggested by SigOpt;
this allows the user to focus on the actual design of the deep learning
model rather than the infrastructure used during hyperparameter tuning.

In this paper, we discuss the circumstances which led us to develop
Orchestrate and the design decisions made to address these
circumstances.
We present an internal use case (conducted during alpha
testing) and goals for future development.

\section{Initial investigation and understanding expectations}
Many organizations face the need to develop scalable infrastructure
to support tools that have been developed in a local environment.
Airflow\footnote{\url{https://airflow.apache.org/}} was developed
at AirBnB to implement and monitor sequences of tasks in a distributed
and asynchronous environment.  
Mesosphere\footnote{\url{https://mesosphere.com}} provides enterprise
solutions around deploying containers to public clouds.
Uber has developed
Michelangelo\footnote{\url{https://eng.uber.com/michelangelo/}} to
provide internal teams the ability to deploy their machine learning
tools at scale.

To inform the development of Orchestrate, 
we interviewed SigOpt users (and, particularly, deep learning users who 
evaluate multiple models in parallel) to understand what was needed for
Orchestrate to be effective.  Below, we summarize the responses
and the resulting Orchestrate workflow.

\subsection{Parallelism\label{sec:parallelism}}
% Customers were interested in leveraging parallel workflows in several ways.
At the highest level, interviewees stated that they want the power to
execute multiple hyperparameter optimization experiments simultaneously
(to accelerate the optimization process); 
furthermore, each of these experiments
could have drastically different compute times.
Within a single experiment, interviewees wanted to be able to leverage
evaluating multiple model configurations simultaneously.  
Even within a single model configuration evaluation, 
interviewees wanted to distribute their model across multiple GPUs and
evaluate multiple cross-validation folds simultaneously.
When initially scoping Orchestrate, we had anticipated the desire
to support evaluating multiple models in parallel, but,
after the interviews, it became clear that for this project
to be successful we would have to address parallelism on multiple levels.

\subsection{On Demand Cluster\label{sec:ondemandcluster}}
Experimental model development may proceed at an erratic pace; interviewees
reported needing significant compute resources at inconsistent intervals
because of the development / tuning cycle at their company.  Additionally,
these interviewees were hoping to leverage the elastic nature of cloud
computing to have access to the resources they needed, when
they needed them.

\subsection{Heterogeneous Resources\label{sec:heterogeneousresources}}
In combining the desire to limit compute cost with a desire to tune
multiple models simultaneously (mentioned in \secref{sec:parallelism}),
we realized that the cluster should be able to support heterogeneous
compute resources.  Both CPU and GPU machines should be available
within the same cluster to
allow models which do not require GPU resources to not have to pay
for them.

%Users often need numerous GPUs per experiment. Our goal was the the user should not be constrained by the machines they own, they should be able to harness the full power of cloud computing. The user can, on demand, spin up a cluster with as many GPUs as they need, and then, again on demand, spin down the cluster. 

\subsection{Monitoring}
In entrusting Orchestrate to manage the infrastructure,
interviewees voiced concern about loss of
proximity to the actual functioning of the model tuning experiment; in our interviews
we learned that they still want to be able to monitor the process
despite abstracting away some of the details.
%Users want to monitor their experiment both during and after the experiment is completed. 

\paragraph{Monitor status} The process of monitoring Orchestrate
status seemed to split into two key factors: the status of the cluster
and the status of individual experiments on the cluster.
Effective monitoring of Orchestrate would require both the ability
to answer the question ``Is the cluster infrastructure operating as
planned?'' and the question ``How is work being distributed for each
of my experiments?''

%keep track of the experiment's progress. Has the model started evaluating any sets of hyperparameters? How many sets of hyperparameters? How many models are running in parallel? Essentially, is my infrastructure holding up? Am I likely to have bugs in my code? How many evaluations are left in this experiment? For clarity, we separate the "status" of an optimization job to be related to infrastructure. We separate this from an experiments' mathematical performance when evaluated on a set of hyperparameters. 

\paragraph{View Logs} A common fear among interviewees was that if the
infrastructure were managed by Orchestrate that they would lose the
ability to easily detect and rectify mistakes in their models.
This is especially complicated in situations where their models
could behave very differently for different model configurations.
The proposed solution
was to be able to easily access logs from experiments
as they were running (or after they crashed).  In particular,
interviewees wanted to be able to easily recover all the logs associated
with a single experiment, irrespective of what other experiments
were running on the cluster or how parallel model configurations were distributed.

%View the log output of every model. Check for warnings / errors in the code. Help debug issues in the model code.

\paragraph{Monitor performance} Perhaps most importantly, interviewees
wanted to be able to monitor the actual quality of the models as
they went through this model tuning process.  Because this was
already managed through the SigOpt website, it was not considered
as a component of Orchestrate.

\subsection{Stopping Experiments}
Interviewees stated that hyperparameter optimization can surface bugs within their 
models, whether due to their code throwing exceptions or their model's performance failing to reach a threshold. 
In both circumstances, interviewees wanted the ability
to terminate all execution on their experiment and free up the
resources for future work.

\subsection{Resulting workflow from our investigation}

% \begin{wrapfigure}[20]{r}[0pt]{0.57\textwidth}
% 	\centering
% 	\vspace{-4.1mm}
% 	\includegraphics[width=.99\linewidth]{orchestrate_workflow}
% 	\vspace{1mm}
% 	\caption{
%         The results of our investigation into a desired workflow for 
%         a tool that could successfully help someone run hyperparameter 
%         optimization experiments.
%         \label{fig:orchestrate_workflow}
%     }
% \end{wrapfigure}

\begin{figure}[ht]
    \centering
    \includegraphics[width=.6\linewidth]{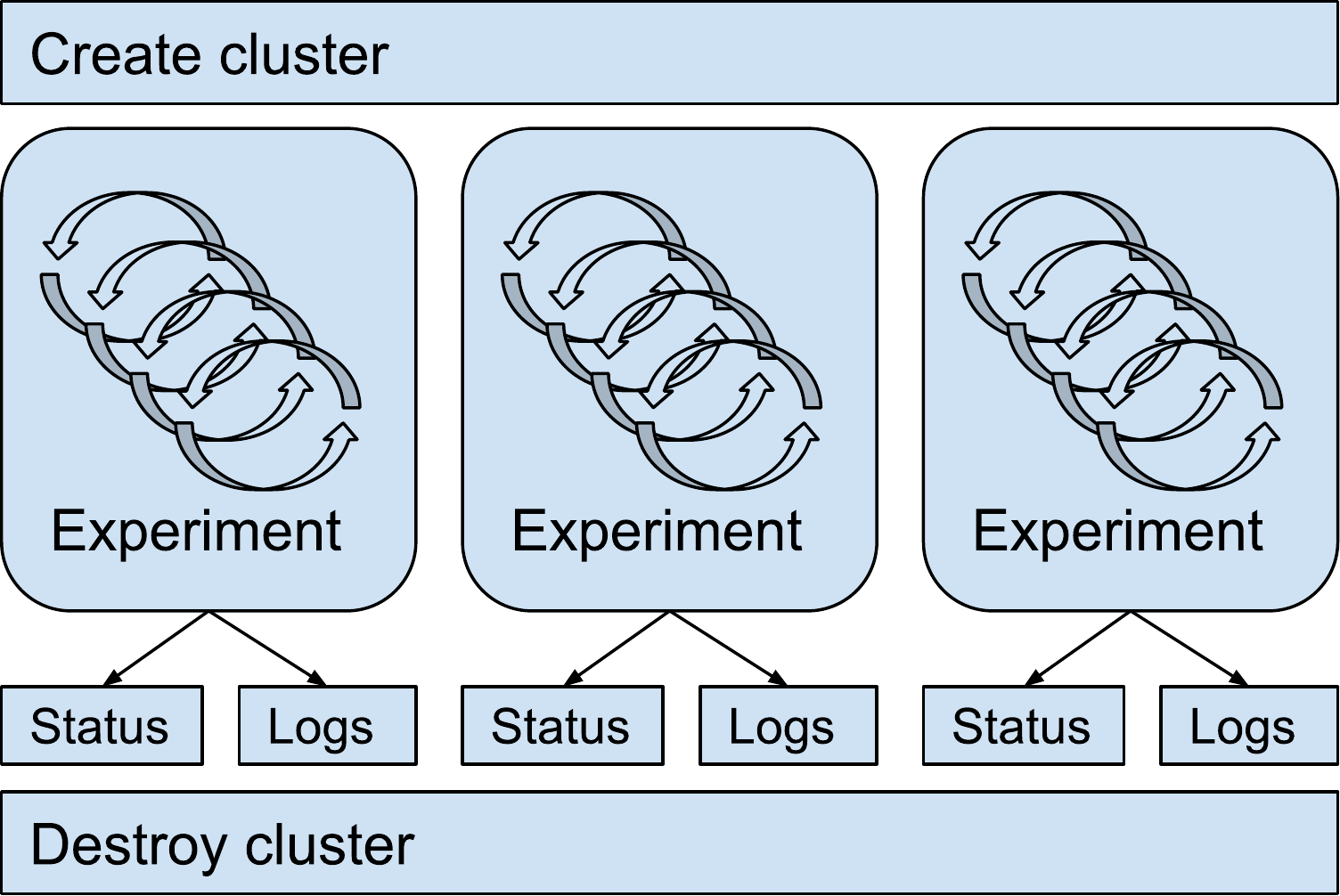}
    \caption{
        The results of our investigation into a workflow to support
        prospective users of Orchestrate with the parallel components of
        their hyperparameter optimization.
        \label{fig:orchestrate_workflow}
    }
\end{figure}

The Orchestrate workflow, as guided by these customer discussions,
is depicted in \figref{fig:orchestrate_workflow}.
Of particular note is that the creation and
destruction of the cluster is dissociated from the running of
experiments.  This allows multiple experiments to be run on a single
cluster (rather than tying the existence of a cluster to a single
experiment) and it allows model tuning artifacts
(such as model-generated logging output)
to remain available after the experiment has completed.
% \begin{enumerate}
%     \item Create a cluster
%     \item Tune several models (simultaneously or sequentially) on this cluster
%     \begin{itemize}
%         \item For each model,
%             run multiple simultaneous SigOpt suggestions 
%             on machines in cluster
%         \item Monitor experiment status, logs, and performance
%         \begin{itemize}
%             \item Based on monitoring, customers can choose to end
%                 experiments early
%         \end{itemize}
%         % \item (Pick One)
%         % \begin{itemize}
%         %     \item  Wait until experiment is completed, or,
%         %     \item Kill experiment before completion
%         % \end{itemize}
%     \end{itemize}
%     \item Destroy the cluster
% \end{enumerate}

\section{Design and Implementation}
% After the initial scoping of the project was completed, we designed and
% implemented a solution.  This section explains the decisions that were
% made and their benefits/drawbacks.

\subsection{Command Line Interface}
Because our customers have a variety of modeling backgrounds and goals,
we prioritized building a model agnostic tool. This tool could be installed anywhere on a user's system, and could be used to containerize models that lived anywhere. We also wanted our tool to be language-agnostic, so that even if our tool is written in Python, it can be used from any environment, on any kind of model.

Our core API commands are listed below;
they, respectively, allow a user to create a cluster, start an experiment, monitor experiment status, view experiment logs, delete an experiment, and destroy an experiment.
\begin{verbatim}
    sigopt cluster create -f cluster_configuration.yml
    sigopt run -f experiment_configuration.yml
    sigopt status $EXPERIMENT_ID
    sigopt logs --follow $EXPERIMENT_ID
    sigopt delete $EXPERIMENT_ID
    sigopt cluster destroy -n $CLUSTER_NAME
\end{verbatim}

% They, respectively, allow a user to create a cluster, start an experiment, monitor experiment status, view experiment logs, delete an experiment, and destroy an experiment.

\subsection{Containerization}
To fulfill our goal of allowing customers to take their
locally developed models and tune them in the cloud, we needed
a mechanism for moving a model (and all its supporting components).
We found containers\footnote{\url{https://www.docker.com/}}
to be a logical solution for managing the process.
This process is less brittle than copying individual files and more
flexible than attempting to provide customers with pre-defined hard disc
images already loaded with standard dependencies.
Orchestrate uses Docker, an industry standard tool for containerization.

One remaining point of contention when using containers is how to move/access
data from which a model should be trained; at present, we advise
users import their data from a cloud source after the container has
started.  Devising a strategy for reducing data traffic is a priority
going forward.

% We thought of scp'ing each file, but that process is brittle and can leave an inconsistent state. We thought of zipping up the users' current directory, with a venv, scp'ing it, and then unzipping and sourcing the venv on the host. This process is brittle because it relies on a particular venv name, and the user may not need all dependencies from their venv in training their model. We also considered using a pre-baked ML AMI that had all packages needed installed. Ultimately, we landed on containers because they are an emerging standard for wrapping up code and dependencies that we had used with some success in previous projects, including a GPU-based ML model blog post. The downside of containers, and of many of these methods, is that we're still left with a question about how to access the data.

\subsection{Kubernetes}
If containers are the emerging standard for wrapping up code and dependencies, Kubernetes\footnote{\url{https://kubernetes.io/}} is the emerging standard for running containers. It is not purpose-built for machine learning (the stated goal of many of our users), but it has many features desired of Orchestrate clusters, such as facilitating communication between machines, starting containers across multiple machines, and managing running containers.
% Kubernetes generally does not suffer from cloud lock-in, you can run a cluster on AWS, Google, and Azure clouds. Indeed, open source tools like kops exist to help you with just that. However, most of these services offer their own, hosted Kubernetes deployments, such as AWS EKS, GKE, and AKS, which speed development of applications such as ours.
Kubernetes provides built-in APIs and abstractions for starting jobs, monitoring infrastructure, and viewing container logs, and other important functions. During our development of Orchestrate we relied heavily on the standard Kubernetes paradigms (such as jobs, pods and containers) and APIs. Using these built-in tools shaved weeks off of our development cycle.

Additionally, managed Kubernetes implementations are hosted by Amazon, Google and Azure clouds. This allowed us to save development time by relying on a managed Kubernetes implementation without being locked-in to one cloud provider.

\subsection{AWS and EKS} 
We chose Amazon Web Services (AWS) as our cloud provider because it was an environment 
that our early interviewees were familiar with and comfortable using. 
AWS released their Elastic Kubernetes Service (EKS) shortly after we began
scoping this project.  To facilitate transfer of containers from local
environments to an EKS cluster, we use AWS Elastic Container Registry (ECR).
Furthermore, EKS allows users to create clusters with both
CPU and GPU machines within the same cluster;
this helps support the ``multiple experiments, one cluster'' goal
described in \secref{sec:ondemandcluster}.

EKS came with a few limitations, however.  EKS is billed separately from AWS machines, so while the cost of one EKS cluster may be negligible compared to a GPU machine, it is still an additional cost.  Additionally, as of this article, AWS limits each account to three EKS instances by default.
To avoid friction from requiring customers contact AWS support to exceed that limit, we opted to build a workflow for a team to share a single cluster to run multiple experiments.

We expect to integration Orchestrate with every other cloud over time so it is fully agnostic to the underlying infrastructure as we progress toward GA.

\subsubsection{Cluster Configuration}

\begin{wrapfigure}[10]{r}[0pt]{0.52\textwidth}
	\centering
	\vspace{-.4in}
\begin{verbatim}
   # Cluster Config file: demo.yml
   cloud_provider: aws
   cluster_name: orchestrate-cluster
   gpu:
     instance_type: p3.8xlarge
     min_nodes: 4
     max_nodes: 4
   cpu:
     instance_type: c4.xlarge
     min_nodes: 4
     max_nodes: 4
\end{verbatim}
    \caption{Example cluster configuration yaml file\label{fig:configfile}}
\end{wrapfigure}

AWS EKS simplifies the process of creating a Kubernetes cluster on AWS, but it still requires knowledge of the AWS and Kubernetes APIs. With Orchestrate, we wanted to go one step further in reducing the complication of spinning up a cluster. When the user creates a cluster, they provide us with a short model configuration yaml file listing the cluster name, cloud provider (only AWS for now), and desired number and type of GPU and / or CPU resources. From this, Orchestrate manages spinning up and connecting the necessary cloud resources behind the scenes.

% \begin{figure}[ht]
%     \begin{verbatim}
%     # Cluster Configuration file: cluster.yml
%     cloud_provider: aws
%     cluster_name: orchestrate-cluster
%     gpu:
%       instance_type: p3.8xlarge
%       min_nodes: 4
%       max_nodes: 4
%     cpu:
%       instance_type: c4.xlarge
%       min_nodes: 4
%       max_nodes: 4
%     \end{verbatim}
%     \vspace{-.3in}
%     \caption{Example cluster configuration yaml file\label{fig:configfile}}
% \end{figure}
% \mnote{consider wrapping}

\subsection{Experiments}
\begin{figure}[ht]
    \centering
    \includegraphics[width=.99\linewidth]{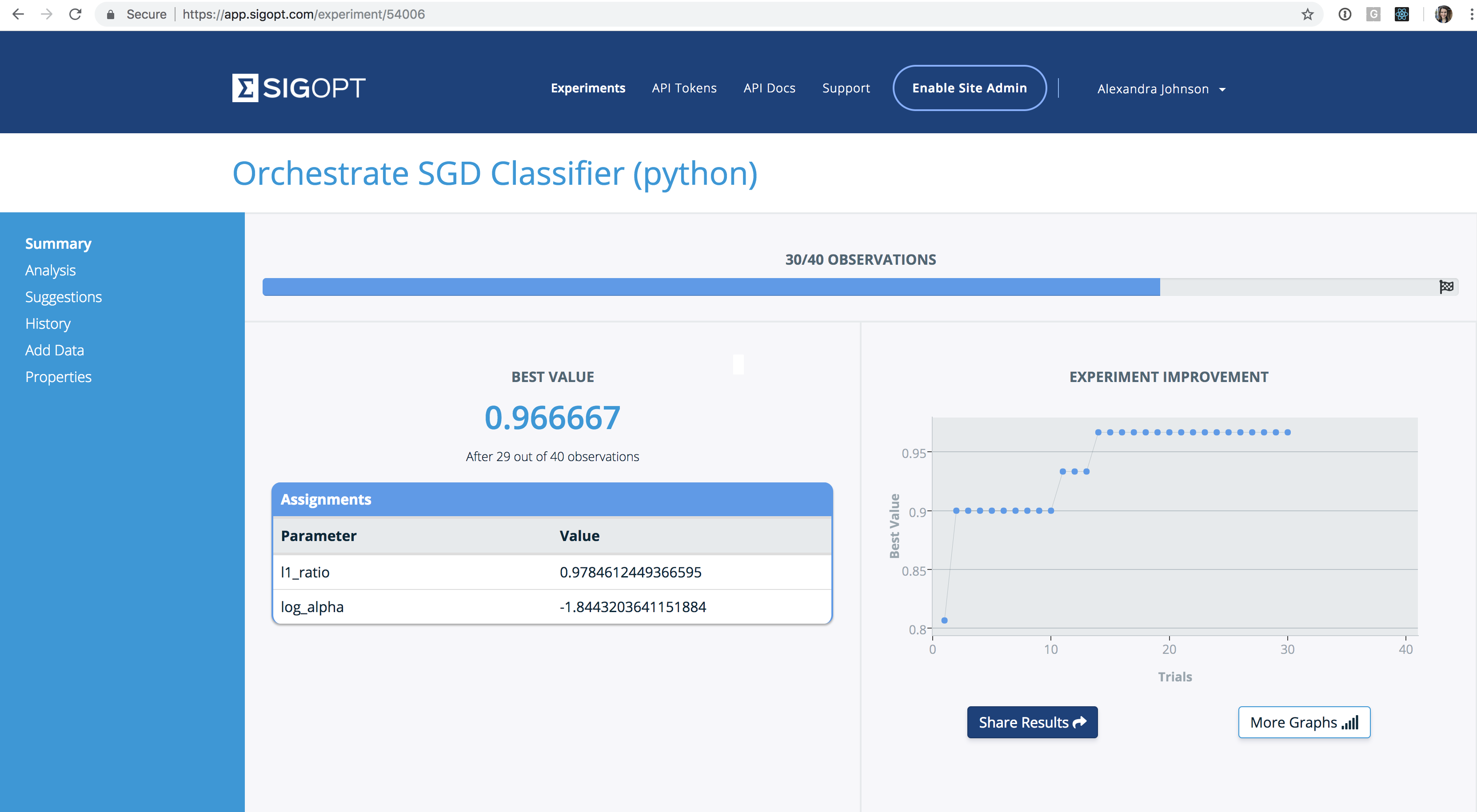}
    \caption{
        An in-progress Orchestrate experiment, viewed
        on \url{https://sigopt.com}.
        \label{fig:running_experiment}
    }
\end{figure}

Orchestrate relies heavily on SigOpt's centralized API for distributed hyperparameter optimization to run model tuning experiments with simultaneous model evaluations.  SigOpt also informs Orchestrate on the progress of the experiment so that experiments can be run for the desired duration (this information can also be recovered from the CLI to monitor experiment status).
Additionally, use of the SigOpt API for experiments allows us to take advantage of the SigOpt web interface to view and share experiments, as shown in \figref{fig:running_experiment}.

SigOpt serves as a system of record for completed experiments. 
While some experiment artifacts, such as container logs, will be lost once
the the cluster is destroyed, experiment metadata, including parameters and performance 
will exist on SigOpt in perpetuity.

\subsubsection{Experiment Configuration}
For experiments requiring GPUs, users may provide the number of GPUs needed per model in their experiment configuration yaml file. Orchestrate passes this information to Kubernetes for use in creating a job on the cluster, and the Kubernetes scheduler manages resource and capacity limitations in the cluster.
The experiment configuration yaml file is where the user defines the experiment structure (number of different configurations with which they wish to evaluate their model and how many of those evaluations may be run in parallel).
\subsection{Limitations}

\paragraph{Models in Development} Because of our focus on model evaluations in parallel, we have not built features and visualizations, etc., that could be useful for a practitioner during early development.

\paragraph{User-provided containers} Because Orchestrate packages a model, dependencies, and Orchestrate-specific code into a Docker container, allowing a user to bring their own model container would run into a technical constraint of running Docker within Docker.

\paragraph{Non-Kubernetes Cluster Management} At present, this tool does not play nicely with in house clusters that are not Kubernetes based. Specifically, we cannot support clusters running Slurm, a popular workload manager for universities. 

\paragraph{High-GPU Models}
The largest GPU instance type provided by AWS currently has 8 GPUs (p3.16xlarge); because Orchestrate currently only supports using AWS, and we have not tested a model exceeding the constraints of one AWS EC2 instance, a single model configuration cannot be trained on more than 8 GPUs simultaneously.

% \subsection{Final structure}

% \paragraph{sigopt cluster create} We use the AWS and the Kubernetes API to create a cluster according to a yaml cluster configuration file.

% \paragraph{run} We use the Docker API to containerize a directory on the user's filesystem, which must contain the user's model and related code. We use the SigOpt API to start an experiment. We configure settings according to a yaml experiment configuration file. The Kubernetes API starts the job. This command prints the experiment id.

% \paragraph{logs} Here, we use the Kubernetes API to view logs for every container running for a given experiment (see \figref{fig:status_and_logs_big}).  By using the ``follow'' argument we can view the logs in real time.

% \paragraph{status} Here, we use the Kubernetes API to find the status of the Kubernetes job for a given experiment. We also use the SigOpt API to identify information about the experiment. We combine these two sources together in the output of the status command.

% \paragraph{delete} (Optional) If you'd like to stop execution on an experiment while it's still running, or simply remove all traces of an experiment from a cluster, the delete command uses the Kubernetes API to delete the Kubernetes job for a given experiment.

% \paragraph{cluster destroy} The user then destroys the cluster, using the name specified in the cluster configuration file. This is done with the AWS API.

\begin{figure}[ht]
    \centering
    \includegraphics[width=.99\linewidth]{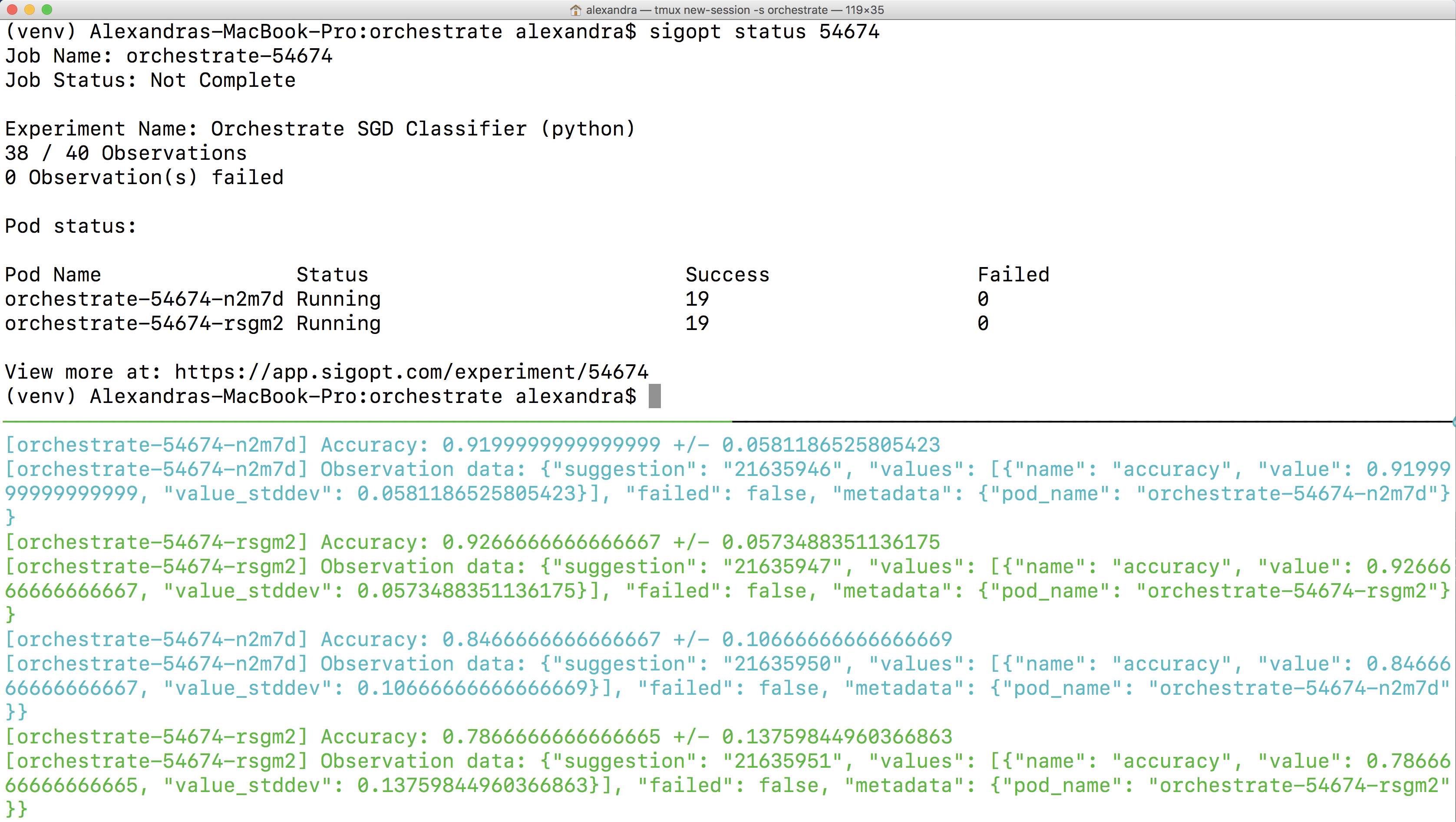}
    \caption{
        A split-screen terminal showing two SigOpt Orchestrate CLI 
        commands. On the top is ``\texttt{sigopt status \$ EXPERIMENT\_{}ID}'',
        and on the bottom is ``\texttt{sigopt logs --follow \$ EXPERIMENT\_{}ID}''.
        The green and blue colors in the logs represent output from
        two distinct simultaneous evaluations of the model.
        \label{fig:status_and_logs_big}
    }
\end{figure}

\section{Initial Feedback}
Our initial alpha tester used Orchestrate over the course of several weeks during his development of a convolutional neural network with 3 convolutional layers and 2 fully connected layers. This neural network was trained on the German traffic sign dataset \cite{stallkamp2011german}, and each model required training on one GPU. During each model tuning experiment with Orchestrate the model was evaluated 300 times, with fifteen models evaluated simultaneously.

The alpha tester said that, in addition to being happy with the numerical results of the hyperparameter optimization, Orchestrate was ``easy to use---I was able to get up and running very fast.'' The constructive feedback aligned with some of our earlier thoughts on Orchestrate's limitations. Specifically, the alpha tester found that Orchestrate was ``a useful tool once you've defined your model but hard if you want to make incremental changes.''  The user also stated that he found the ability to extract logs during and after the run to be ``helpful''.

\section{Future Work}
High priority areas for future work include:
 
\begin{itemize}
    \item Incorporating information about how much storage or computational resources each model requires.
    \item Efficient dataset storage on the cluster. Good strategies for managing this have been discussed at the NIPS ML Systems workshop \cite{sung2017nsml}.
    \item Connecting to existing Kubernetes clusters not created by Orchestrate.
    \item Supporting other cloud providers.
    \item Meet new collaborators to understand the needs of specific use
    cases, e.g., reinforcement learning or natural language processing.
\end{itemize}

\bibliographystyle{plain}
%\bibliography{citations}

\end{document}